\documentclass[twocolumn,aps,amssymb,showpacs]{revtex4}

\usepackage{graphicx}
\usepackage{dcolumn}
\usepackage{bm}
\begin{document}


\title{Comparison of near-interface traps in Al$_2$O$_3$/4H-SiC and Al$_2$O$_3$/SiO$_2$/4H-SiC structures}

\author{Marc Avice}
\author{Ulrike Grossner}
\email{ulrike.grossner@fys.uio.no}
\author{Ioana Pintilie}
\author{Bengt G. Svensson}%
\affiliation{%
University of Oslo, Physics Department and Center for Materials Science \& Nanotechnology (SMN)\\
P.O. Box 1048, Blindern, N-0316 Oslo, Norway
}%
\author{Ola Nilsen}%
\author{Helmer Fjellvag}%
\affiliation{%
University of Oslo, Chemistry Department and Center for Materials Science \& Nanotechnology (SMN)\\
P.O. Box 1033, Blindern, N-0315 Oslo, Norway
}%

\date{\today}

\begin{abstract}
Aluminum oxide (Al$_2$O$_3$) has been grown by atomic layer
deposition on n-type 4H-SiC with and without a thin silicon
dioxide (SiO$_2$) intermediate layer. By means of Capacitance
Voltage and Thermal Dielectric Relaxation Current measurements, the interface
properties have been investigated. Whereas for the samples with an
interfacial SiO$_2$ layer the highest near-interface trap density is
found at 0.3~eV below the conduction band edge, E$_c$, the samples
with only the Al$_2$O$_3$ dielectric exhibit a nearly trap free
region close to E$_c$. For the Al$_2$O$_3$/SiC interface, the
highest trap density appears between 0.4 to 0.6~eV below E$_c$. The
results indicate the possibility for SiC-based
MOSFETs with Al$_2$O$_3$ as the gate dielectric layer in future high
performance devices.
\end{abstract}
\maketitle
\clearpage
\newpage

With the development of the semiconductor research and industry in recent years, a significant interest in advanced materials for volume production of diode and transistor devices has arised.
One of the most promising of these materials is SiC due to its superior chemical and thermal inertness and high electrical break-down field.
Furthermore, because of its wide band gap of 3.26~eV for the most stable polytype 4H, a higher information transfer density for broadcast applications is possible~\cite{cmp04}.
Another interesting area is the utilization of SiC devices in hybrid electric vehicles because of the reduction in the size, weight, and cost of the power conditioning and thermal systems compared to conventional ones~\cite{tolbert:765}.\\
\indent Unfortunately, structures utilizing SiC's natural oxide, SiO$_2$, as a dielectric suffer from a density of shallow interface states below the the conduction band edge, E$_c$, at least two orders of magnitude higher than for comparable Si-based devices. 
These electron traps are suggested to be 'near-interface traps' and attributed to intrinsic defects in the interfacial region of SiO$_2$~\cite{afa:321,pensl:3.2,afa:336}.
Furthermore, the electron channel mobility in 4H-SiC metal-oxide-semiconductor field-effect transistors (MOSFET) is reduced by at least one order of magnitude relative to the bulk mobility~\cite{olaf}. 
Apart from this, SiO$_2$ has a quite low dielectric constant and is thermally not as stable as SiC. With a higher dielectric constant, thinner gate oxides may be used in coherence with the demand for smaller structures.
Therefore, initial efforts have been made in recent years to investigate alternative gate oxides on SiC with a particular objective to minimize the density of states, D$_{it}$, close to E$_c$.
One of the most promising candidates is aluminum oxide, Al$_2$O$_3$, with a reported dielectric constant of $\approx$10, a large band gap ($\approx$6.2~eV), a good thermal stability, and reasonably large conduction ($\approx$1.7~eV) and valence band offsets ($\approx$1.2~eV) to 4H-SiC~\cite{afa:1839}.\\
\indent Within this study, the interface traps in Al$_2$O$_3$/SiC metal-oxide-semiconductor (MOS) devices are directly measured by means of the Thermal Dielectric Relaxation Current (TDRC) technique.
Furthermore, a comparison of the electrical properties, based on TDRC measurements, of the SiO$_2$/SiC and Al$_2$O$_3$/SiC interfaces is made.
It will be shown, that the near-interface trap density close to E$_c$ is higher for Al$_2$O$_3$/SiO$_2$/SiC than for Al$_2$O$_3$/SiC capacitors.
This indicates either that dangling bonds at the interface are saturated in the Al$_2$O$_3$/SiC system or that the responsible traps rather appear within the SiO$_2$ layer than at the interface to SiC, which is also suggested by recent theoretical findings~\cite{knaup:115323}.\\
\indent A 100nm thick Al$_2$O$_3$ layer has been grown by Atomic Layer Chemical Vapour Deposition (ALCVD) on Si-faced, n-type 4H-SiC samples with a 10~$\mu$m thick epi-layer (doping level 2$\times$10$^{15}$~cm$^{-3}$) on a highly doped substrate (1$\times$10$^{18}$~cm$^{-3}$), oriented 8$^\circ$ off the (0001) direction, purchased from Cree Inc, following the surface cleaning and growth procedure reported in Ref.~\cite{avice:xxx}. 
To obtain the intermediate SiO$_2$ layer in the second set of samples, SiC has been dry-oxidized at 1150~$^\circ$C for 10~min before the aluminum oxide growth, resulting in an SiO$_2$ layer of approximately 5~nm in thickness.
Before depositing circular Al contacts (diameter 0.5~mm) by thermal evaporation through a shadow mask, the Al$_2$O$_3$/SiC samples have been annealed in argon for 2~h at 1100~$^\circ$C, resulting in crystallization of the Al$_2$O$_3$ layer~\cite{avice:jap}.
No thermal treatment has been applied to the Al$_2$O$_3$/SiO$_2$/SiC samples. 
Silver paste has been used as an Ohmic back side contact.
The samples have been characterized by TDRC measurements in the temperature range between 40 and 320~K and Capacitance Voltage (CV) measurements at room temperature using a probe frequency of 1~MHz and a sweep rate of 0.5~V/s.\\
\begin{figure}[t]
\centering
\vspace{-1.3cm}
\includegraphics[scale=0.76]{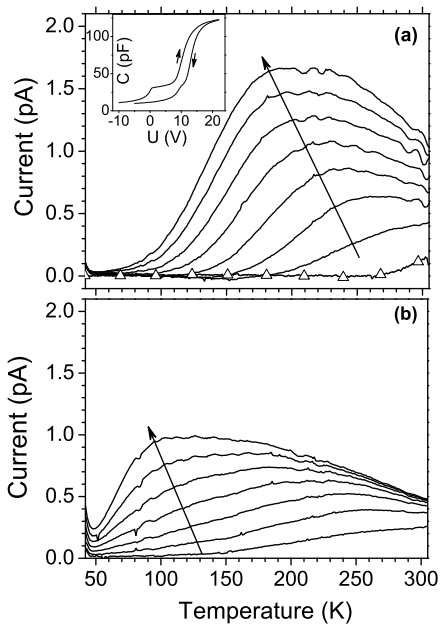}
\vspace{-0.5cm}
\caption{\label{cv} TDRC measurements of an Al/Al$_2$O$_3$/SiC capacitor using a charging temperature of (a) 330~K and (b) 40~K for a discharging voltage V$_{discharging}$=-5V and a heating rate $\beta$=0.133~K/s. V$_{charging}$ was varied in steps of 2~V from 6 to 20~V in (a) and from 6 to 18~V in (b). The inset in (a) shows the CV characteristics at room temperature, the curve indicated with triangles the current during heating under short-circuit conditions.}
\end{figure}
\indent TDRC and CV measurements of the Al/Al$_2$O$_3$/SiC capacitors are shown in Fig.~\ref{cv}.
As presented in the inset of panel (a), the Al/Al$_2$O$_3$/SiC capacitors exhibit a flatband voltage (V$_{FB}$) of about 11~V.
The plot is obtained by sweeping from deep depletion to accumulation and backwards. 
The small capacitance step at about 0V is found in almost any Al$_2$O$_3$/SiC capacitor, and mobile ions introduced during growth are a possible origin~\cite{avice:jap}.
In general, the CV characteristics are similar to those reported recently for Al$_2$O$_3$/SiC samples~\cite{avice:xxx}. \\
To obtain the TDRC spectra, the SiC surface is brought into accumulation by applying a forward bias (V$_{charging}$) at elevated temperature, or, alternatively, low temperature.
After cooling to 40~K under forward bias, a reverse bias (V$_{discharging}$) is applied in order to place the capacitor into deep depletion.
The temperature is subsequently raised at a constant rate $\beta$, and filled traps in the upper part of the bandgap begin to emit electrons to the SiC conduction band edge, E$_c$.
Hence, an emission current is observed due to the electrons being swept out of the depletion region, as shown in the data sets in Fig.~\ref{cv}. It should be noted that the field within the MOS structure is sufficiently large to sweep out the carriers out of the depletion region without any recombination and, hence, causing a current which is only due to the electrons emitted from the traps~\cite{blood}.
The TDRC measurements have been performed for different charging voltages, keeping the discharging voltage (V$_{discharging}$=-5V) and heating rate ($\beta$=0.133~K/s) constant.
The corresponding leakage current (the current measured for V$_{discharging}$ when no filling of the traps is performed - not shown in the figure) was substracted from the recorded TDRC spectrum.
To verify whether there is an overlying current due to dipole polarization/depolarization to the emptying of the traps within the TDRC measurements, the samples were cooled under V$_{discharging}$ and the short-circuit current was measured during the heating (curve indicated with triangles in Fig.~\ref{cv}(a)). 
It has been found to be around zero for temperatures T$\leq$270K. 
In the upper panel (a), a charging temperature of 330~K was used, whereas in the lower panel, (b), the sample was charged at 40~K (duration 15 s).
\begin{figure}[t]
\suppressfloats
\centering
\vspace{-0.55cm}
\includegraphics[scale=0.8]{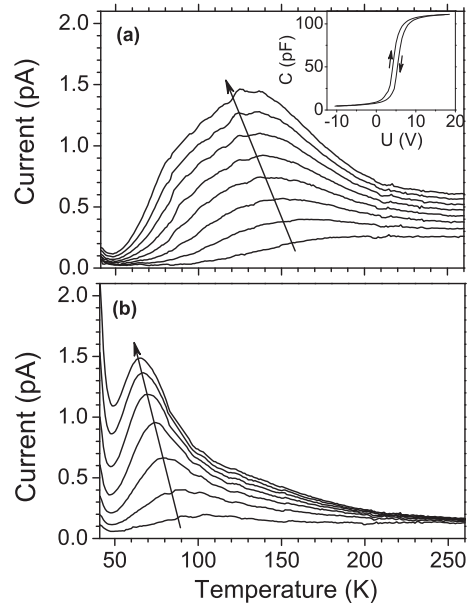}
\vspace{-0.3cm}
\caption{\label{ct} TDRC measurements of an Al/Al$_2$O$_3$/SiO$_2$/SiC capacitor using a charging temperature of (a) 345~K and (b) 40~K for a discharging voltage V$_{discharging}$=-2V and a heating rate $\beta$=0.133~K/s. V$_{charging}$ was varied in steps of 2~V from 6 to 20~V in (a) and from 6 to 18~V in (b). The inset in (a) shows the CV characteristics at room temperature.}
\end{figure}
In (a), a broad peak between 150 and 250~K develops with increasing charging voltage, but no signal is found in the low-temperature range of the spectrum, indicating a low density of shallow electron traps. 
For a charging temperature of 40~K, a different peak at about 100~K develops with increasing charging voltage, whereas the peak found in Fig.~\ref{cv}(a) is strongly suppressed.
The substantial increase of the broad peak in Fig.~\ref{cv}(a) with charging voltage (accumulation) demonstrates that it originates from near-interface traps and is not caused by traps in the SiC bulk.
Moreover, the strong suppression of this peak in Fig.~\ref{cv}(b) reflects presumably a pronounced temperature dependence of the electron capture cross section and/or a spatial location of the traps which extends into the Al$_2$O$_3$ layer yielding a thermally activated filling process.
Also the peak at $\approx$100~K in Fig.~\ref{cv}(b) increases in intensity with charging voltage and thus, the corresponding traps occur close to the interface.
Further, these traps are filled at low temperature (40~K) showing an electron capture cross section which does not vanish at low temperatures and a depth distribution mainly confined to the interface with minor penetration into the Al$_2$O$_3$ layer.
The absence of a TDRC signal below $\approx$100~K when starting the charging at 330~K, Fig.~\ref{cv}(a), may indicate that the 100~K peak in Fig.~\ref{cv}(b) is due to a bi- or metastable defect, i.e., the atomic configuration with minimum energy and the associated trap levels depend on the defect charge state during cooling.\\
\indent In Fig.~\ref{ct}, TDRC spectra for the Al$_2$O$_3$/SiO$_2$/SiC capacitors are presented using different charging voltages at 345~K (a) and 40~K (b).
Two different groups of traps are revealed with peak positions at $\approx$140~K and 70~K, respectively, and the amplitude of both groups increases with the charging (accumulation) voltage, showing that they are located at (or close to) the SiO$_2$/SiC interface.
When charging at 40~K, the 70~K peak dominates and the broad one at $\approx$140~K is reduced.
In fact, the TDRC spectra in Fig.~\ref{ct} and their dependence on charging voltage and charging temperature display close resemblance with that recently reported for SiO$_2$/4H-SiC capacitors~\cite{rudenko:545}.
Hence, the properties of the Al$_2$O$_3$/SiO$_2$/SiC capacitors appear to be dominated by the SiO$_2$/SiC interface.
Rudenko et al.~\cite{rudenko:545} have argued that both groups of traps are due to the same type of intrinsic interfacial defect of acceptor character with a spatial distribution extending from the interface into an oxycarbide transition region.
\begin{figure}[t]
\centering
\vspace{-0.8cm}
\includegraphics[scale=0.73]{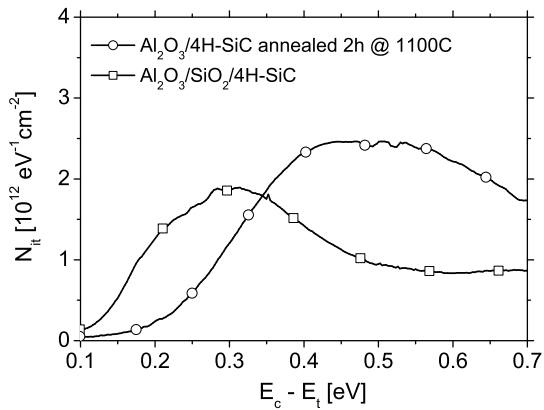}
\vspace{-0.5cm}
\caption{\label{ed} Energy distribution of the near-interface traps for an Al/Al$_2$O$_3$/SiC capacitor (data set with circles) and an Al/Al$_2$O$_3$/SiO$_2$/SiC capacitor (data set with squares). The energy scale refers to the conduction band of 4H-SiC.}
\vspace{-0.5cm}
\end{figure}
The 70~K peak is assigned to communication of the defect trap with the SiC conduction band edge while the broad peak at $\approx$140~K is ascribed to communication with the conduction band edge of the oxycarbide transition layer, exhibiting a gradually increasing offset ranging from the conduction band edge of 4H-SiC to that of SiO$_2$.
Indeed, in light of recent theoretical findings the defect may be identified as a pair of carbon atoms substituting for oxygen (C$_O$=C$_O$), which is a stable center giving rise to an electron trap within SiO$_2$ with a position close to the conduction band edge of 4H-SiC~\cite{knaup:115323}.\\
Following the procedure outlined by Simmons and Mar~\cite{simmons:3865,mar:131}, the trap energy distribution has been deduced from the TDRC data in Figs.\ref{cv} and \ref{ct}, and the results for charging at 330 (345)~K are presented in Fig.~\ref{ed}. 
In this context it should be emphasized that the prime objective of the present study is not to minimize the absolute values of D$_{it}$ but rather to compare the D$_{it}$ versus energy distribution for the Al$_2$O$_3$/4H-SiC and Al$_2$O$_3$/SiO$_2$/4H-SiC capacitors.
Figure \ref{ed} reveals clearly that the former ones are essentially free of electron traps for energies $\le$0.2~eV below E$_c$ while the latter ones exhibit a high D$_{it}$ close to E$_c$, in accordance with previous reports for the SiO$_2$/4H-SiC interface~\cite{afa:321,pensl:3.2,afa:336,rudenko:545}.
On the other hand, the Al$_2$O$_3$/4H-SiC samples display a high density of deep states from $\approx$0.4 to $\approx$0.7~eV below E$_c$ and these are due to the traps with a thermally activated filling process, as discussed in conjunction with Fig.~\ref{cv}.\\
\indent In conclusion, Fig.~\ref{ed} shows unambiguously that the intrinsic and shallow near-interface traps dominating in SiO$_2$/4H-SiC structures do not appear in Al$_2$O$_3$/4H-SiC capacitors and at least two possible explanations can be put forward; these intrinsic defects do not form at the Al$_2$O$_3$/4H-SiC interface, which may be consistent with the assignment to a C$_O$=C$_O$ pair~\cite{knaup:115323}, or they are efficiently passivated. In any case, it can be concluded that Al$_2$O$_3$ shows great promise as gate dielectric for 4H-SiC MOSFET's with a low density of shallow interface states which limit the electron channel mobility and further work is being pursued with a particular emphasis to minimize the density of deep states between $\approx$0.4 and $\approx$0.7~eV below E$_c$ and to reduce the flat band voltage.

Financial support by the Norwegian Research Council (Strategic University Program) and the University of Oslo (FUNMAT program) is greatly acknowledged.


\bibliography{nittsc}

\end{document}